# Robust H-infinity Adaptive Fuzzy Approach for Unknown Nonlinear Networked Systems

Li Jun Heng, Wang Yong Weiwei

*Abstract*—An H-infinity adaptive fuzzy control design is proposed in this paper for unknown nonlinear networked systems. The main issues of networked systems are addressed here; which are the system delay (may cause system instability) and loss of information. In fact, the proposed method overcomes the delays by filtering the errors and also compensate the loss of system information. The adaptive fuzzy control design is combined in this work with H-infinity control approach to approximate the system's unknown nonlinear functions. The stability analysis of the approach is also surveyed through Lyapunov theory. The results revealed that the closed loop system stability is proved in existence of system disturbances, system delays and information loss. The proposed approach is applied on an inverted pendulum system to evaluate the method's efficiency and effectiveness.

## I. Introduction

The network technologies are being widely useful recently since they are cheap and easy to implement. Some instances of networked systems that are implemented in the control systems are wide area network (WAN), local area network (LAN), and Internet. The mentioned networked systems are used in the control structure in order to exchange useful information between control system through a network [1]. The networked systems are highly applicable in industrial processes such as robots control, intelligent transportation system, WSN (Wireless Sensor Networks), power systems, web-based control systems, etc.

There are several issues associated with the networked control systems (NCS)s, such as data quantization issue, security issue, information loss, system delay, etc [2]. However, the two prominent reasons of system degradation are the data loss, and the network delay. Hence, the networked systems control strategies have been under certain attention in the recent years. The regular control strategies are designed based on the non-delayed actuations assumptions, so they need to be modified before applying on the networked systems.

Based on the control theory, the unknown nonlinear system can be first converted to a linear system and then the adaptive fuzzy approach can be used to approximate the system model. Various research works implemented different fuzzy approaches for system approximation, for instance [3] used the sliding mode adaptive fuzzy approach for delay compensation in NCSs. Also, [4] used an LMI (Linear Matrix Inequalities) method to build the state feedback controllers to model the networked control system as a system with delayed inputs, the researcher also studied the stability aspect of the control system. The studies in [4-6] designed the control strategy offline with kind of LMI approaches to compensate the network delay and system information loss.

Regarding that the nonlinear networked systems are subject to system uncertainties and disturbances, the fuzzy adaptive approaches seem to be appropriate for system approximation. In [7], the author implemented the indirect adaptive fuzzy approach on an inverted pendulum with uncertainties; the approach also showed to maintain the closed-loop stability. Other researchers benefited from the sliding mode techniques as the robust methods to the uncertain networked systems. However, the sliding mode control approaches bother from the system measured noises and chattering [8-11].

Thus, in this paper a new approach to deal with the two main issues of network delay and system information loss associated with the networked systems is introduced. In this control method, the adaptive fuzzy controller is coupled with the H-infinity auxiliary control to avoid disturbances of modeling error. Through this approach, the output signal converges and the closed-loop control signal is bounded and stable. The robust H-infinity loop shaping approach to decrease the disturbances in the flexible beam system is used before by [12], however in this work it is tried to use the H-infinity mixed with the adaptive fuzzy method to avoid delay and the data loss in the networked system.

The paper is organized as follows. First, the problem formulation and its parameters are represented in detail in section II. Section III describes the H-infinity adaptive fuzzy controller as the proposed approach. Section IV gives the results of the approach on inverted pendulum and discusses about them. Eventually, the paper is concluded in section V.

## II. Problem Formulation and Mathematical Modeling

Consider the following state space formulation (1) as the nonlinear system descriptive model. Note that any nonlinear system can be represented as (1) by the feedback linearization method.

$$\dot{x}_1 = x_2$$
$$\dot{x}_2 = x_3$$
$$\vdots$$
$$\dot{x}_n = f(x_1, \cdots, x_n) + g(x_1, \cdots, x_{n2})u(t) + d(t)$$
$$y = x_1 \qquad (1)$$

In the above formulation, $f$ and $g$ are nonlinear functions. Moreover, $y$ and $u$ are the output and the output of the system respectively. $x$ is the state vector of the system and $d(t)$ is the disturbance signal.

Considering the delay in the network, the system (1) can be represented as (2).

$$\dot{x}_1 = x_2$$
$$\dot{x}_2 = x_3$$
$$\vdots$$
$$\dot{x}_n = f(x_1, \cdots, x_n) + g(x_1, \cdots, x_{n2})u(t-\tau) + d(t)$$
$$y = x_1 \quad (2)$$

$\tau$ is the network delay in the system. Considering that the desired trajectory is $x_d$, the objective is that the system states track the desired path in existence of network delay, disturbances, and unknown nonlinear functions of the system formulation.

Thus, the error of tracking the desired trajectory would be as $e = x_d - x$.

### III. H-INFINITY ADAPTIVE FUZZY CONTROL DESIGN

The robust H-infinity control design structure is illustrated in Fig. 1 [12].

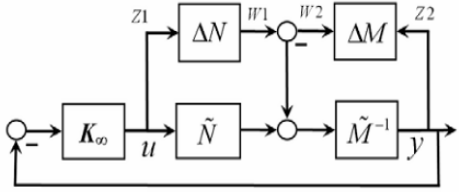

Fig. 1: H-infinity control design structure

Regarding the above figure, the nominal model of the system is chosen as $P$, and the pre and post compensators are chosen as W1 and W2 respectively. The structure of the plant using the H-infinity control would be as (3).

$$Ps = W2 * P * W1 = \begin{bmatrix} As & Bs \\ Cs & Ds \end{bmatrix} \quad (3)$$

where matrices $As, Bs, Cs, Ds$ are the linearized state space representation of the H-infinity controlled plant.
If there exist a controller $K_\infty$ such that equation (4) is obtained, then the closed-loop plant Ps with H-infinity controller is stable [12].

$$\|Tzw\|_\infty = \left\| \begin{bmatrix} I \\ K_\infty \end{bmatrix} (I + PsK_\infty)^{-1} \right\| \leq \frac{1}{\varepsilon} \quad (4)$$

By minimizing $\frac{1}{\varepsilon}$, the robustness of the system would increase.

The indirect adaptive fuzzy control design for estimation of the system model is chosen from [7]. The fuzzy structure is based on the Mamdani fuzzy system; the objective is to build a feedback controller with adaption law $u(x,\theta)$ such that the system output $y$ tracks the desired trajectory by regulating the parameters vector $\theta$.

So, the unknown nonlinear functions $f$ and $g$ need to be estimated as $\hat{f}$ and $\hat{g}$ through the if-then fuzzy rules which are attained from the system input-output behaviors. Moreover, there are some free parameters in between that can be tuned to improve the estimation process.

Then, the $u$ signal can be stated as (5).

$$\hat{f}(x) = f(x,\theta)$$
$$\hat{g}(x) = g(x,\theta) \quad (5)$$

Based on the mentioned mathematical models, the block diagram of the adaptive fuzzy control strategy is shown is Fig. 2.

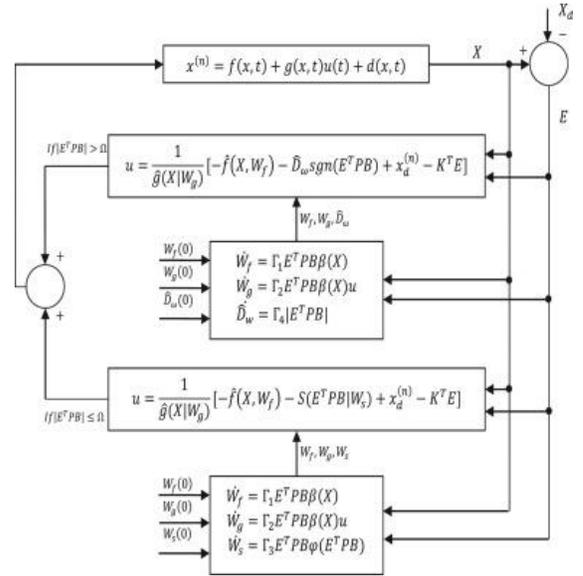

Fig. 2: Adaptive fuzzy control structure

### IV. SIMULATION RESULTS

The fuzzy control toolbox of MATLAB is used to implement the fuzzy adaption law. For evaluating the approach effectiveness, the control design is applied on an inverted pendulum benchmark problem [7]. The inverted pendulum is chosen as the plant because it is high order, unstable, and highly coupled. Furthermore, this system is very useful in robots, flight control and so on.

The simulations are run with the existence of system disturbances. The output of the system is attained with the objective of tracking the desired signal. The reference signal is chosen as a sine wave. Also, the tracking error is illustrated in the continuation.

The plant output tracking trajectory (including the output signal and the reference trajectory), and the tracking error trajectory are shown in the following figures.

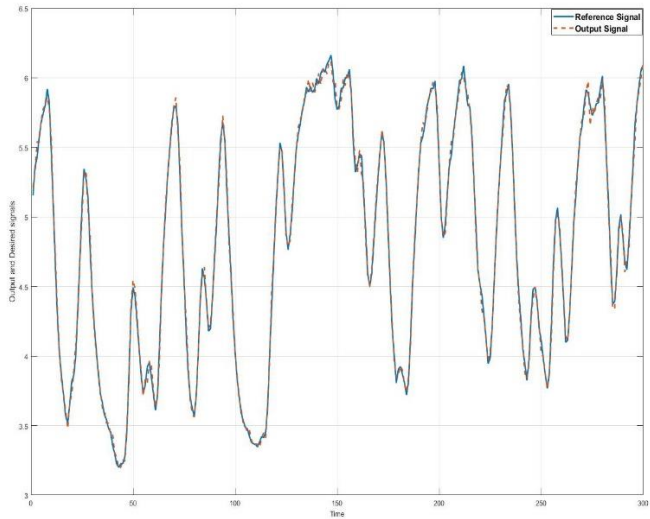

Fig. 3: Output signal and reference trajectory of inverted pendulum system, using H-infinity adaptive fuzzy controller

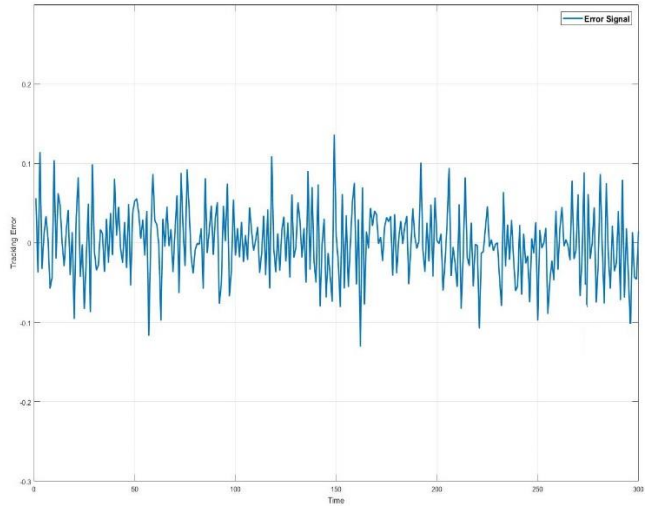

Fig. 4: Tracking error signal of inverted pendulum system, using H-infinity adaptive fuzzy controller

From Fig. 3, it is proved that the reference tracking is perfectly performed through the proposed H-infinity adaptive fuzzy approach. Also the error signal in Fig. 4 admits that the steady state value of error doesn't go more than 10 percent (fluctuates around zero) and it is a satisfactory result for the inverted pendulum system. Moreover, the stability of the closed-loop system is attained; the system is asymptotically stable at zero. Additionally, the controller totally overcomes the system disturbances as it is clearly shown in the figures.

Therefore, the pendulum system is estimated well using the proposed control method in existence of couplings and networked framework.

## V. CONCLUSIONS

The H-infinity adaptive fuzzy control approach is developed for a networked unknown nonlinear system. The model disturbances are also considered in the system. The objective is to avoid system delay due to the network structure, disturbances and the loss of information.

To evaluate the proposed approach, the method is implemented on the inverted pendulum benchmark problem since this system is very applicable and common in industrial applications.

According to the results, the objectives are met. The reference signal tracking is perfectly done with error signal less than 10 percent. Also, the error signal stabilizes eventually with the steady state value.

For future work, implementing the proposed approach on a networked chaotic benchmark problem is recommended.

REFERENCES

[1] R. A. Gupta and M. Y. Chow, "Networked control system: Overview and research trends," *IEEE Trans. Ind. Electron.*, vol. 57, no. 7, pp. 2527–2535, Jul. 2010.
[2] *Bemporad, M. Johansson, and M. Heemels,* Networked Control Systems. *Berlin, Germany: Springer, 2010.*
[3] M. A. Khanesar, O. Kaynak, S. Yin, and H. Gao, "Adaptive indirect fuzzy sliding mode controller for networked control systems subject to time-varying network-induced time delay," *IEEE Trans. Fuzzy Syst.*, vol. 23, no. 1, pp. 205–214, Feb. 2015.
[4] M. Yu, L. Wang, T. Chu, and F. Hao, "Stabilization of networked control systems with data packet dropout and transmission delays: Continuous time case," *Eur. J. Control*, vol. 11, no. 1, pp. 40–49, 2005.
[5] J. Sun and J.-P. Jiang, "Stability of uncertain networked control systems," *Proc. Eng.*, vol. 24, pp. 551–557, 2011.
[6] L.-S. Hu, T. Bai, P. Shi, and Z. Wu, "Sampled-data control of networked linear control systems," *Automatica*, vol. 43, no. 5, pp. 903–911, May 2007.
[7] R. Eini and S. Abdelwahed, "Indirect Adaptive fuzzy Controller Design for a Rotational Inverted Pendulum," *2018 Annual American Control Conference (ACC)*, Milwaukee, WI, USA, 2018, pp. 1677-1682. doi: 10.23919/ACC.2018.8431796.
[8] V. Utkin, J. Guldner, J. Shi, S. Ge, F. Lewis, Sliding Mode Control in Electro-Mechanical Systems. Boca Raton: CRC Press, 2009.
[9] C. Edwards, S. Spurgeon, Sliding Mode Control. London: CRC Press, 1998.
[10] S. Ryvkin, E. Palomar Lever, Sliding Mode Control for Synchronous Electric Drives. London: CRC Press, 2011.
[11] E. Slotine and W. Li, Applied Nonlinear Control, Englewood Cliffs, NJ:Prentice-Hall, 1991.
[12] R. Eini, Flexible Beam Robust H-infinity Loop Shaping Controller Design Using Particle Swarm Optimization. Journal of Advances in Computer Research, 5(3), Quarterly pISSN: 2345-606x eISSN: 2345-6078, 2014.